\documentclass[conference]{IEEEtran}
\IEEEoverridecommandlockouts
\usepackage{cite}
\usepackage{amsmath,amssymb,amsfonts}
\usepackage{algorithmic}
\usepackage{graphicx}
\usepackage{textcomp}
\usepackage{pbalance}
\usepackage{color, soul}
\usepackage{xcolor}
\def\BibTeX{{\rm B\kern-.05em{\sc i\kern-.025em b}\kern-.08em
    T\kern-.1667em\lower.7ex\hbox{E}\kern-.125emX}}
\begin{document}

\title{Design and Evaluation of a DC Microgrid Testbed for DER Integration and Power Management}






%

\author{
\IEEEauthorblockN{\textbf{Gokul Krishnan S},  \textbf{Charalambos Konstantinou}}

\IEEEauthorblockA{
CEMSE Division, King Abdullah University of Science and Technology (KAUST)
}

\IEEEauthorblockA{
E-mail: \{gokulkrishnan.sivakumar,  charalambos.konstantinou\}@kaust.edu.sa}

}

\maketitle

\begin{abstract}
This paper presents a DC microgrid testbed setup that consists of various Distributed Energy Resources (DERs) including solar Photovoltaics (PV), supercapacitors for voltage regulation, and Battery Energy Storage Systems (BESS). 
The DC microgrid accommodates both non-flexible and flexible loads which can be dynamically adjusted based on PV power availability. The integration of the setup with the Hyphae Autonomous Power Interchange System (APIS) framework automates energy transfer within the BESS, ensuring efficient power management and optimizing the overall efficiency of the DC microgrid. Furthermore, the setup is validated in terms of the efficacy of the proposed model via real-time simulation, facilitated by the Speedgoat baseline real-time target Hardware-in-the-Loop (HIL) machine. The results demonstrate the model's adeptness in efficiently managing power sharing, emphasizing the capabilities of the DC microgrid setup in terms of performance and reliability in dynamic energy scenarios as well as enhancing the resilience of the grid amidst PV uncertainties. 
\end{abstract}

\begin{IEEEkeywords}
Distributed energy resources, battery energy storage systems, flexible loads, autonomous power interchange system, hardware-in-the-loop. 
\end{IEEEkeywords}

\section{Introduction}
In the past few decades, the growing demand for sustainable energy generation and efficient power distribution has made the shift towards a decentralized energy infrastructure. The integration of Distributed Energy Resources (DER) especially in localized and self-contained energy systems, i.e., microgrids, can help boost the resilience and efficiency of traditional power grids by enabling localized energy generation and distribution, thereby reducing reliance on centralized systems and facilitating the integration of Renewable Energy Sources (RES). Particularly, DC microgrids offer the advantage of improved energy efficiency by eliminating the need for multiple conversions between DC and AC power, which reduces energy losses. This is particularly beneficial for integrating DC-based RES Distributed Generation (DG), like Photovoltaic solar panels (PV) and Battery Energy Storage Systems (BESS), leading to a more efficient and sustainable energy system. Additionally, DC microgrids eliminate concerns related to the management of reactive power flow and frequency regulation, enhancing their operational efficiency and reliability \cite{1, batmani2022sdre}. 
Overall, the shift towards decentralized DC microgrids leads to a highly dynamic platform in which DERs seamlessly integrate into the existing grid infrastructure. 



DC bus voltage regulation and load power sharing are the two main control functions of a DC microgrid. Voltage regulation ensures that there are no deviations in the voltage from the setpoint in the steady-state operation. Power sharing control ensures that each load in the grid gets adequate power for  proper functionality. It also administrates the power flow within the DGs and BESS \cite{2}. In \cite{3}, advanced control strategies such as droop control and hierarchical control systems are implemented to manage the dynamic behavior of the microgrid and maintain grid stability. In \cite{4}, the authors proposed a hybrid adaptive fuzzy-based power management control system to regulate the power-sharing between BESS and supercapacitors.

Numerous studies in the literature have focused on the development of real-time testbed enviroments \cite{katuri2023experimental, zografopoulos2021cyber} and of DC microgrid configurations. In \cite{5}, an experimental DC microgrid testbed setup with Controller-Hardware-in-the-Loop (CHIL) and Power-Hardware-in-the-Loop (PHIL) that considers the dynamics in RES was designed. An adaptive Proportional and Integral (PI) controller was used to control the bidirectional converters to improve the transient as well as steady-state performance of the power flow. In \cite{6}, a 20kW DC microgrid testbed was built to simulate the various grid fault conditions that include DC bus voltage oscillations and utility grid failures. The authors state that the setup can emulate blackout conditions due to the sudden increments ofloads. In \cite{7}, a simulation tool for DC microgrid testing and benchmarking was proposed with the inclusion of various simulators for RES, bidirectional grid power flow, BESS, diesel generators and controllable inductive, capacitive, and electronic resistive loads. The work can validate DC microgrid project development in each stage. In \cite{8}, a centralized energy management system was designed to manage peak and off-peak load demands by optimal load shedding and allocation of PVs and BESS. In [9], it is shown that centralized systems of the type discussed herein are vulnerable to single-point failures and exhibit limited reliability.

In this paper, we design a DC microgrid testbed which incorporates a Speedgoat baseline real-time target HIL machine for obtaining real-time results. The proposed setup includes PV generation, BESS, and supercapacitors for voltage regulation. It also includes non-flexible and flexible dynamic loads to address uncertainties in solar power generation like irradiation change due to partial shading caused by clouds, weather changes, temperature changes, etc. The modeling of those components \cite{10}, and an energy management system has been designed using MATLAB. The Hyphae framework has been used as a decentralized system that manages the power-sharing between different BESS within the DC microgrid \cite{12}. Hyphae is running on a Linux system which in turn communicates with the HIL machine using transmission control protocol/internet protocol (TCP/IP). {Our work integrates the  physical peer-to-peer energy exchange framework, Hyphae, with the DC microgrid model, which consists of different types of sources and loads, and runs in a real-time simulator.} The complete DC microgrid system, encompassing both control and communication subsystems, was designed using Simulink and implemented on a HIL machine.

The rest of the paper is as follows. The design and configuration of the testbed setup including the mathematical modeling of system components are presented in Section \ref{s:testbed}. Real-time results that are obtained from the HIL machine have been analyzed in Section \ref{s:results}. Finally, a conclusion of this work is presented in Section \ref{s:conclusions}.

\section{Testbed Setup: Modeling, Control, and\\ System Integration}\label{s:testbed}

\begin{figure}[t]
\centering
\includegraphics[width=3.3in]{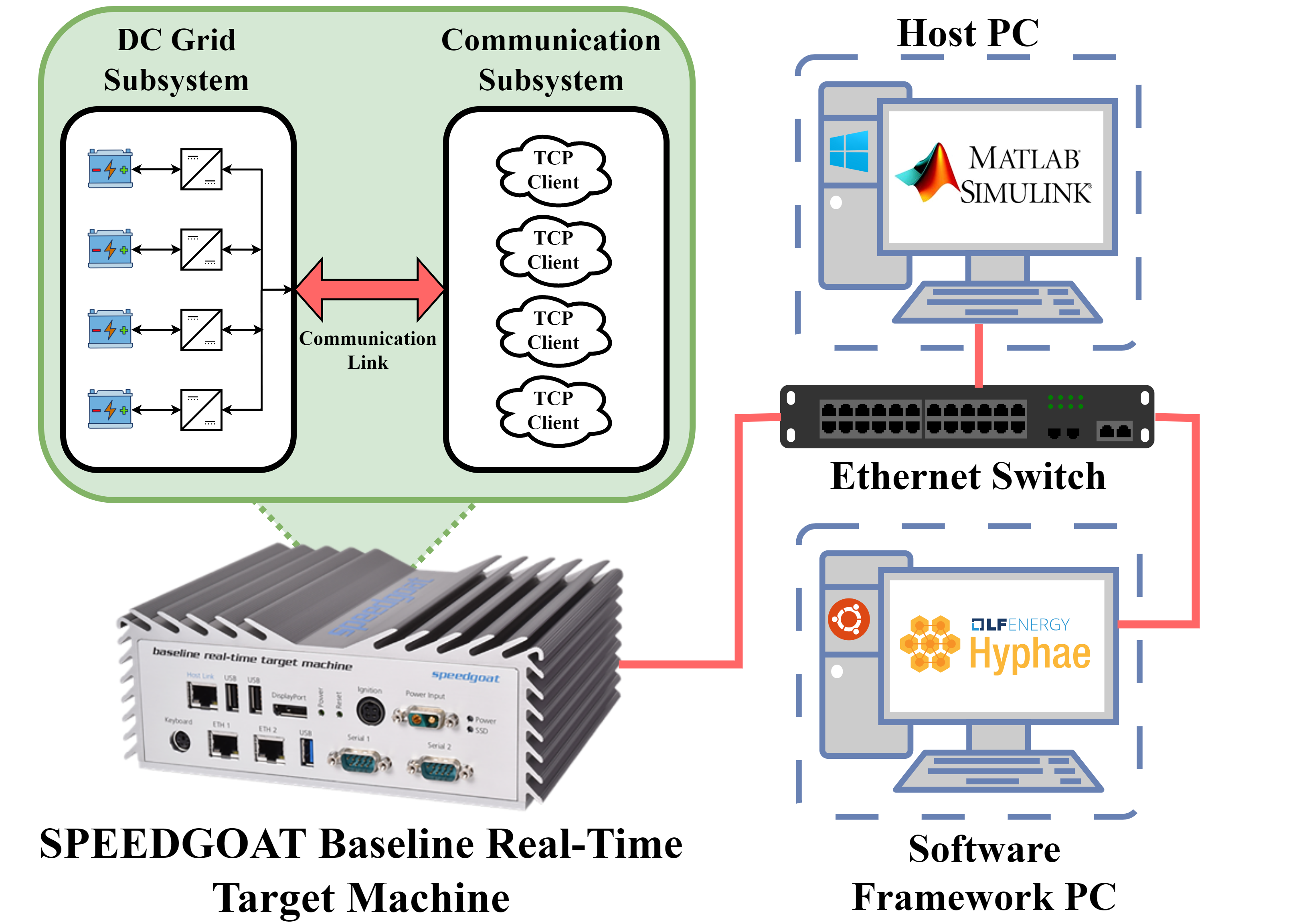}
\caption{Block diagram of the testbed setup with the HIL machine.}
\label{fig_1}
\end{figure}

The experimental testbed setup consists of a high-performance Windows machine running the DC microgrid model and relevant  communication systems within MATLAB/Simulink. Hyphae {Autonomous Power Interchange System (APIS)} software framework is installed and configured to run on an Ubuntu 20.04 LTS Linux distribution. A Speedgoat baseline real-time target HIL machine with the specifications listed in Table \ref{tab1} runs the DC microgrid model with its control and communications with the Hyphae framework through TCP/IP protocol. All the above three machines are connected to the same network using a D-Link gigabit ethernet switch. The entire testbed setup configuration is shown in Fig. \ref{fig_1}. A Python program with sockets has been developed to fetch and write the data from the APIS emulator module  and HIL machine where the DC microgrid model runs in real-time. The HIL machine which runs the DC microgrid model is configured as a TCP server with four different ports for four different nodes in the model. The Python program in the Linux machine acts as a TCP client. The message is encoded as a 6-bit width which receives the deals information and the State-of-Charge (SoC) of each BESS is encoded in ASCII and transferred to Hyphae. The testbed is designed in such a way that all the model parameters like PV irradiation, temperature, SoCs of BESS, and load resistance/power can be varied in real-time when the simulation is running.

\begin{table}[t]
\caption{Target hardware specifications.}
\begin{center}
\begin{tabular}{||c|c||}
\hline \hline
\textbf{Description}&{\textbf{Value}} \\
\hline
CPU& Intel Celeron 2 GHz, 4 cores \\
\hline
Memory (RAM)& 4 GB DDR3 \\
\hline
Main Drive& 64 GB SSD \\
\hline
PCIe& IO191, IO397, IO791 \\
\hline \hline
\end{tabular}
\label{tab1}
\end{center}
\end{table}

\begin{figure}[t]
\centering
\includegraphics[width=3.2in]{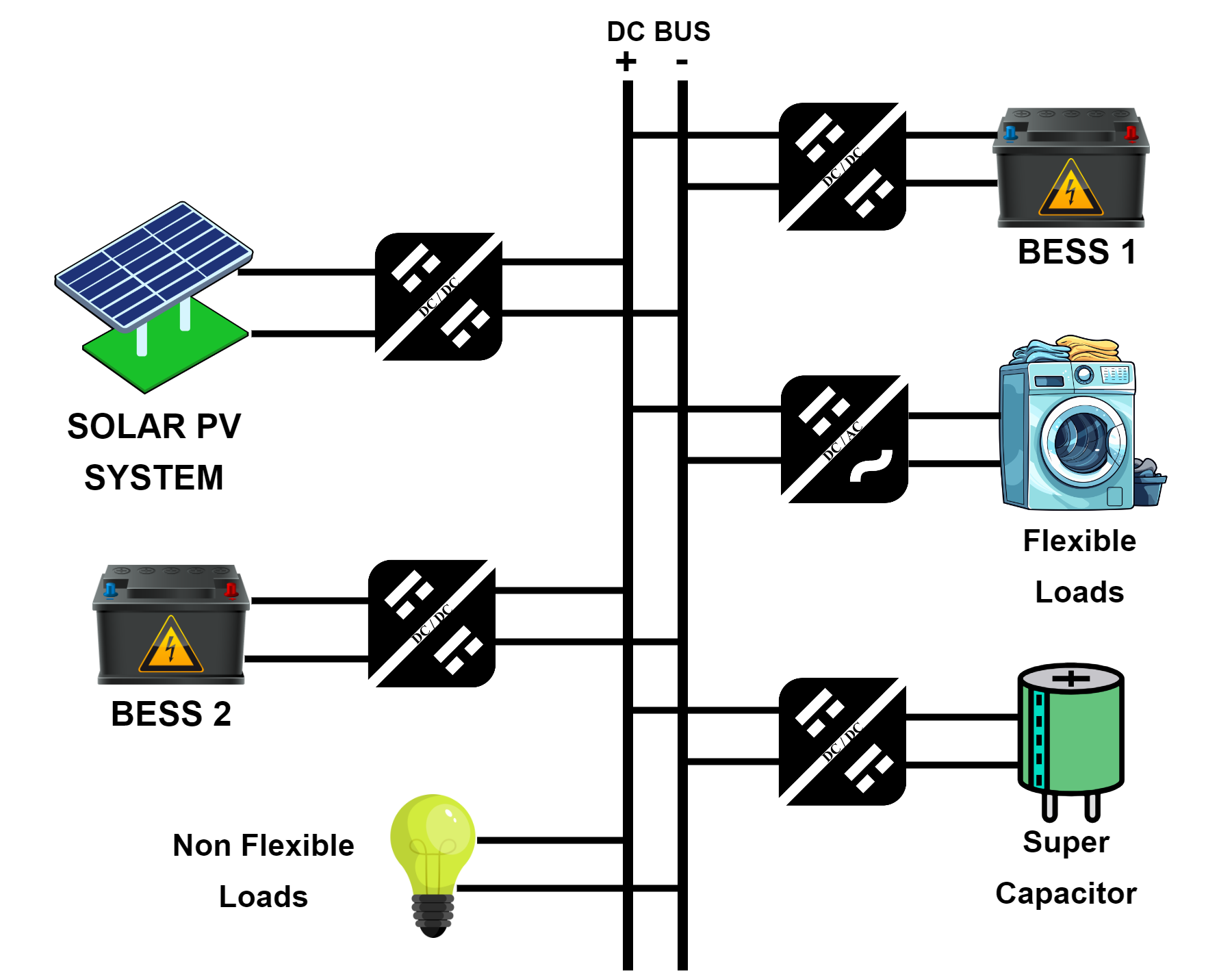}
\caption{DC microgrid model with DERs, non-flexible and flexible loads.}
\label{fig_2}
\end{figure}

A systematic approach for the design and simulation of the DC microgrid, which includes PV arrays, BESS, and supercapacitors, as shown in Fig. \ref{fig_2}, is performed using MATLAB/Simulink. PV, BESS and supercapacitor modeling, non-flexible and flexible dynamic load modeling and bidirectional DC-DC converter design are discussed below.

\subsection{PV System Design and Modeling}
\begin{figure}[t]
\centering
\includegraphics[width=3in]{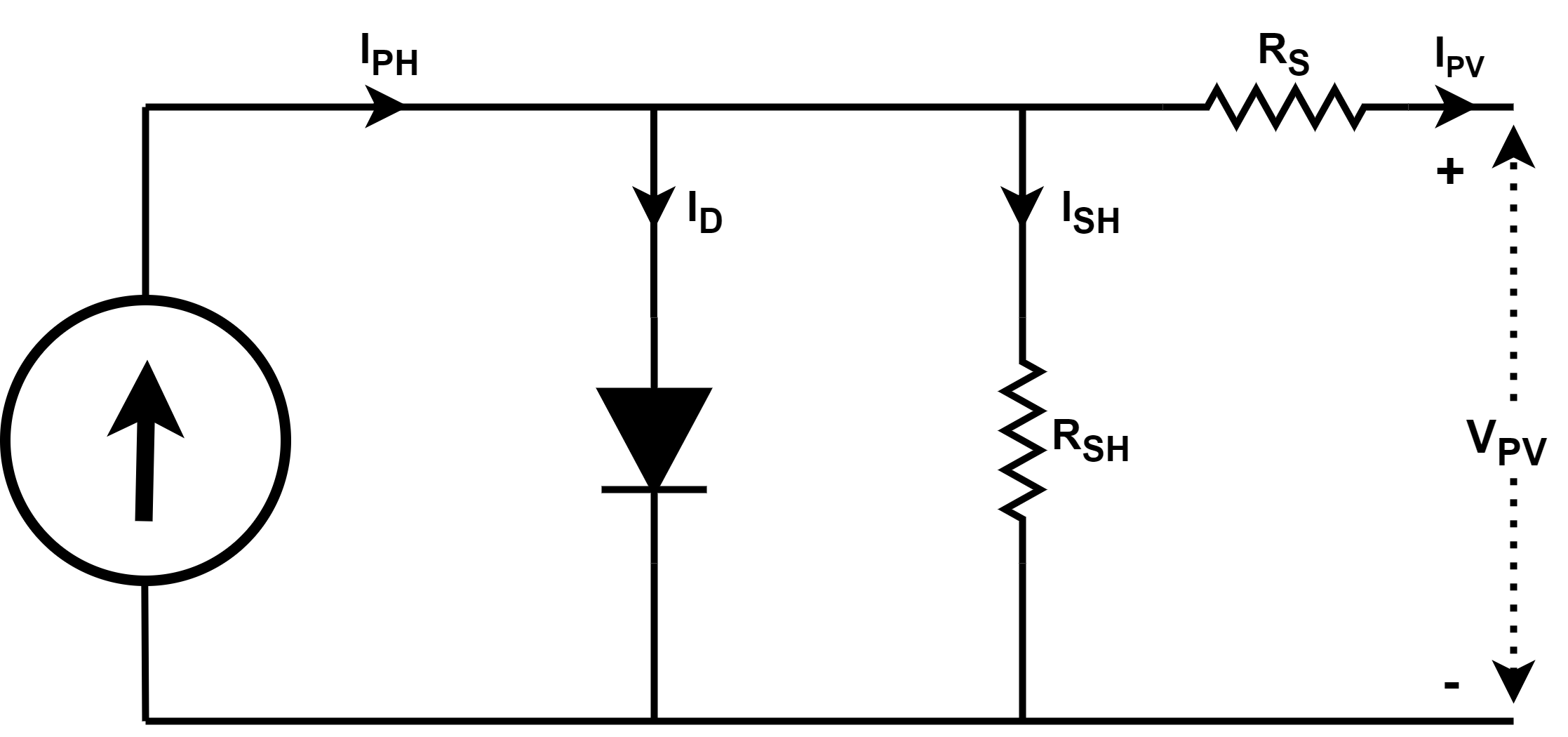}
\caption{Solar PV cell circuit as constant current source.}
\label{fig_4}
\end{figure}

The PV panel can be represented as a constant source with some parametric losses as shown in the circuit of Fig. \ref{fig_4}. Solar PV system works only when there is an adequate amount of solar radiation in the field. On the other hand, it is highly susceptible to uncertainties due to shading and temperature changes. Eqs. \eqref{eq1}--\eqref{eq3} represent the current generated from the solar PV panel with the environmental constraints as input \cite{13}. The PV array block, featuring the Sharp ND-167U1 industry-grade PV module in a 3s10p configuration with a total rated power of 5 kW at 69 V, is designed in Simulink. This design incorporates the perturb and observe method for Maximum Power Point Tracking (MPPT).

\begin{equation}
I_{PV} = GA_{cell}\eta-I_D-I_{SH} \label{eq1}
\end{equation}
\begin{equation}
I_D = I_o(e^{\frac{qv}{nKT}}-1) \label{eq2}
\end{equation}
\begin{equation}
I_{SH} = \frac{V_j}{R_{SH}} \label{eq3}
\end{equation}
\noindent where $I_{PV}$ is the solar cell output current, $G$ is the solar irradiation at the PV panel, $A_{cell}$ is the effective cell area, $\eta$ is the efficiency of the solar cell, $I_D$ is the cell diode current, and $I_{SH}$ is the cell shunt current. In Eq. \ref{eq2}, $I_o$ is the reverse saturation current, $q$ is the elementary charge, $v$ is the voltage across PV cell, $n$ is the ideality factor, $K$ is the Boltzman's constant, and $T$ is the temperature in Kelvin. In Eq. \ref{eq3}, $V_j$ is the voltage across shunt resistor and $R_{SH}$ is the resistance of shunt element in the PV cell.

\subsection{BESS Modeling}
The battery used in the DC grid as a storage element typically has a very high capacity. BESS provides the system with extra resiliency whenever there is an unavailability or uncertainty in the power generating sources.  A battery pack can have different chemistries like Li-Ion, Lead-Acid, Ni-MH, Ni-Cd. It can be represented as a constant voltage source with the requirement of a bidirectional converter to charge and discharge through the grid. The voltage of the battery is represented in Eq. \eqref{eq4} in terms of the battery’s grid current \cite{14}:

\begin{equation}
V_B = E-IR_i \label{eq4}
\end{equation}
\noindent where $V_B$ is voltage of the battery, $E$ is the open circuit voltage of the battery, $I$ is the current through the battery, and $R_i$ is the internal resistance of the battery. 

\subsection{Supercapacitor Modeling}
A supercapacitor has a very high capacitance value and power density compared to batteries. It can discharge its power in seconds. This feature is used to control and regulate the DC grid voltage during the transients, sudden uncertainties in power generating sources as well as load changes. This is achieved by constantly monitoring the DC grid voltage and a sudden fluctuation in this causes a current controller to activate the supercapacitor and regulate the grid voltage. Supercapacitor is also represented as a voltage source as given in Eq. \eqref{eq5} \cite{16}:

\begin{equation}
V_c(t) = [\frac{1}{C_{SC}}\int_{-\infty}^tI(t)dt]-R_{SC}I(t) \label{eq5}
\end{equation}
\noindent where $V_c(t)$ is voltage of the supercapacitor, $C_{SC}$ is capacitance of the supercapacitor, $I(t)$ is the current through the supercapacitor, and $R_{SC}$ is the equivalent series resistance of the supercapacitor.

\subsection{Bidirectional DC-DC Converter Design}

\begin{figure}[t]
\centering
\includegraphics[width=3in]{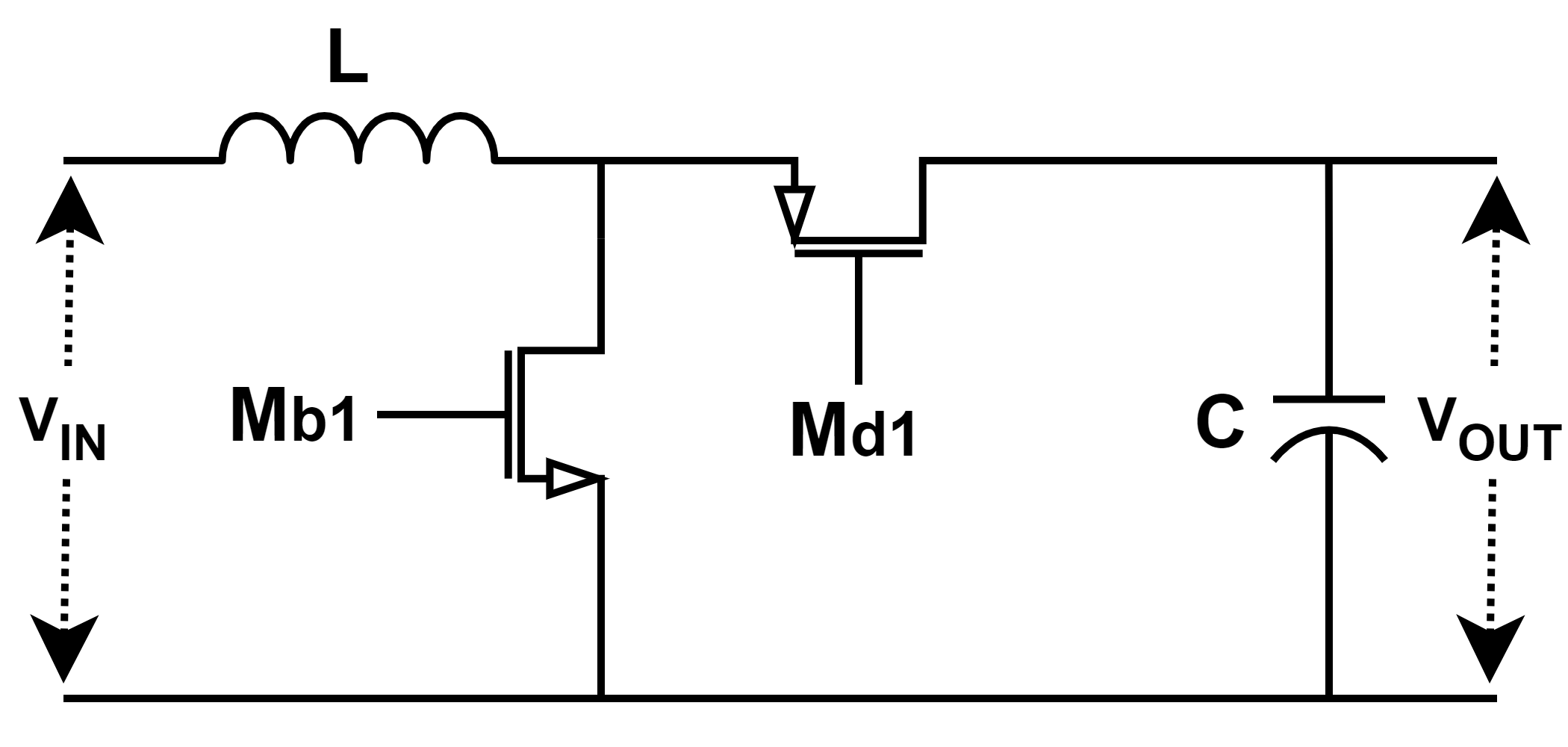}
\caption{Synchronous boost converter circuit.}
\label{fig_5}
\end{figure}

The energy exchange between the BESS, loads, and supercapacitor occurs exclusively through this power electronic converter. Inspired by the conventional boost converter design, this version replaces the diode with a transistor, enabling two-way energy transfer as a synchronous boost converter, as illustrated in Fig. \ref{fig_5}. It operates in two modes: the first is a boost mode, and the second is a current-controlled semi-buck mode. The design specifications include a switching frequency $f$ = 1kHz, input voltage $V_{in}$ = 69V, output voltage $V_{out}$ = 100V, and output current $I_{out}$ = 50Amp, with the converter rated for 5 kW. The sizing of the filter circuits is performed according to Eqs. \eqref{eq6}--\eqref{eq9} \cite{bidcdc}:

\begin{equation}
\triangle I_L = 0.01 \times I_{out} \times \frac{V{out}}{V{in}} = 0.72704Amp\label{eq6}
\end{equation}
\begin{equation}
\triangle V_o = 0.01 \times V_{out} = 1V \label{eq7}
\end{equation}
\begin{equation}
L=\frac{V_{in}(V_{out}-V_{in})}{\triangle I_L f V_{out}} \times 1.5 = 0.044205H \label{eq8}
\end{equation}
\begin{equation}
C=\frac{I_{out}(1-\frac{V_{in}}{V_{out}})}{f \triangle V_o} = 0.01558F \label{eq9}
\end{equation}
\noindent where $\triangle I_L$ is the current ripple, $\triangle V_o$ is the voltage ripple, and $L$ and $C$ are the inductance and capacitance of the converter, respectively.

\subsection{Load Modeling}
The proposed DC microgrid model consists of both non-flexible fixed load, which is modeled as Eq. \eqref{eq10}, and flexible dynamic load, modeled as Eq. \eqref{eq11}. A separate control system is designed to dynamically adjust the flexible load’s power consumption based on the PV power availability which is calculated from Eq. \eqref{eq1} and user-determined constraint input $\gamma$ as the power limiter. The total power consumption of the load is given in Eq. \eqref{eq12} \cite{18}: 

\begin{equation}
P_{non-flex}=V_gI_{nflx} \label{eq10}
\end{equation}
\begin{equation}
P_{flex} = I_{PV}V_g-\gamma \label{eq11}
\end{equation}
\begin{equation}
P_{total}=\sum_{i=1}^NP_i \label{eq12}
\end{equation}
\noindent where $P_{non-flex}$ is the load power consumed, $V_g$ is voltage of the grid, $I_{nflx}$ is the current consumed by the load, $P_{flex}$ is the power consumed by flexible loads, and $P_{total}$ is the total power including non-flexible and flexible loads. 

\subsection{Control Design}
\begin{figure}[t]
\centering
\includegraphics[width=3in]{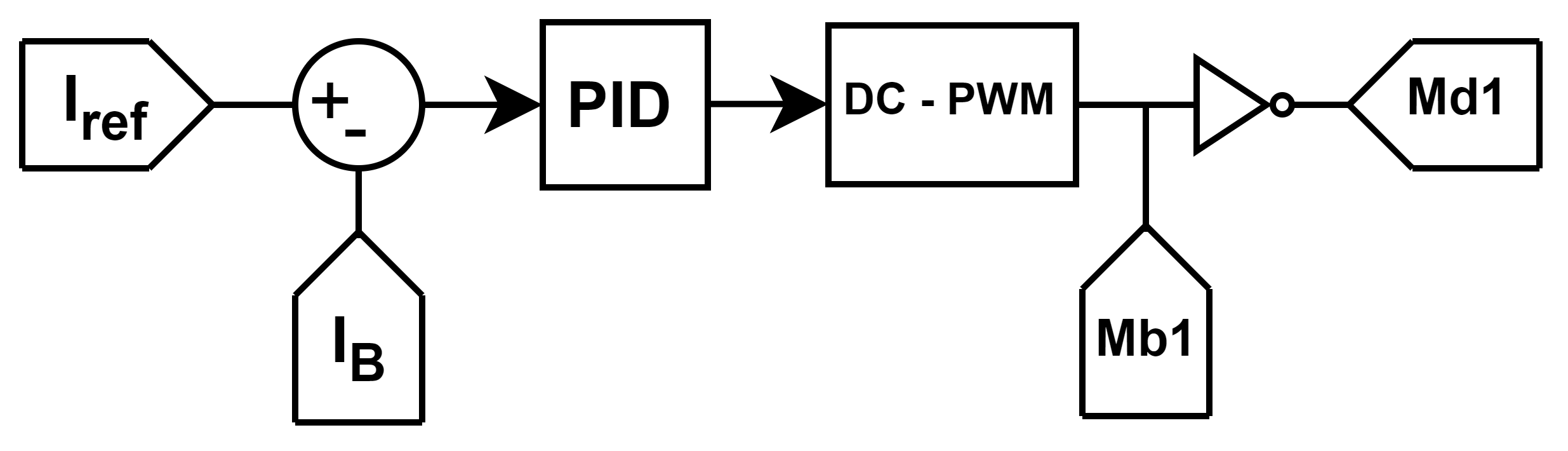}
\caption{PI controller design for DC-DC converter.}
\label{fig_6}
\end{figure}

\begin{figure}[t]
\centering
\includegraphics[width=3.3in]{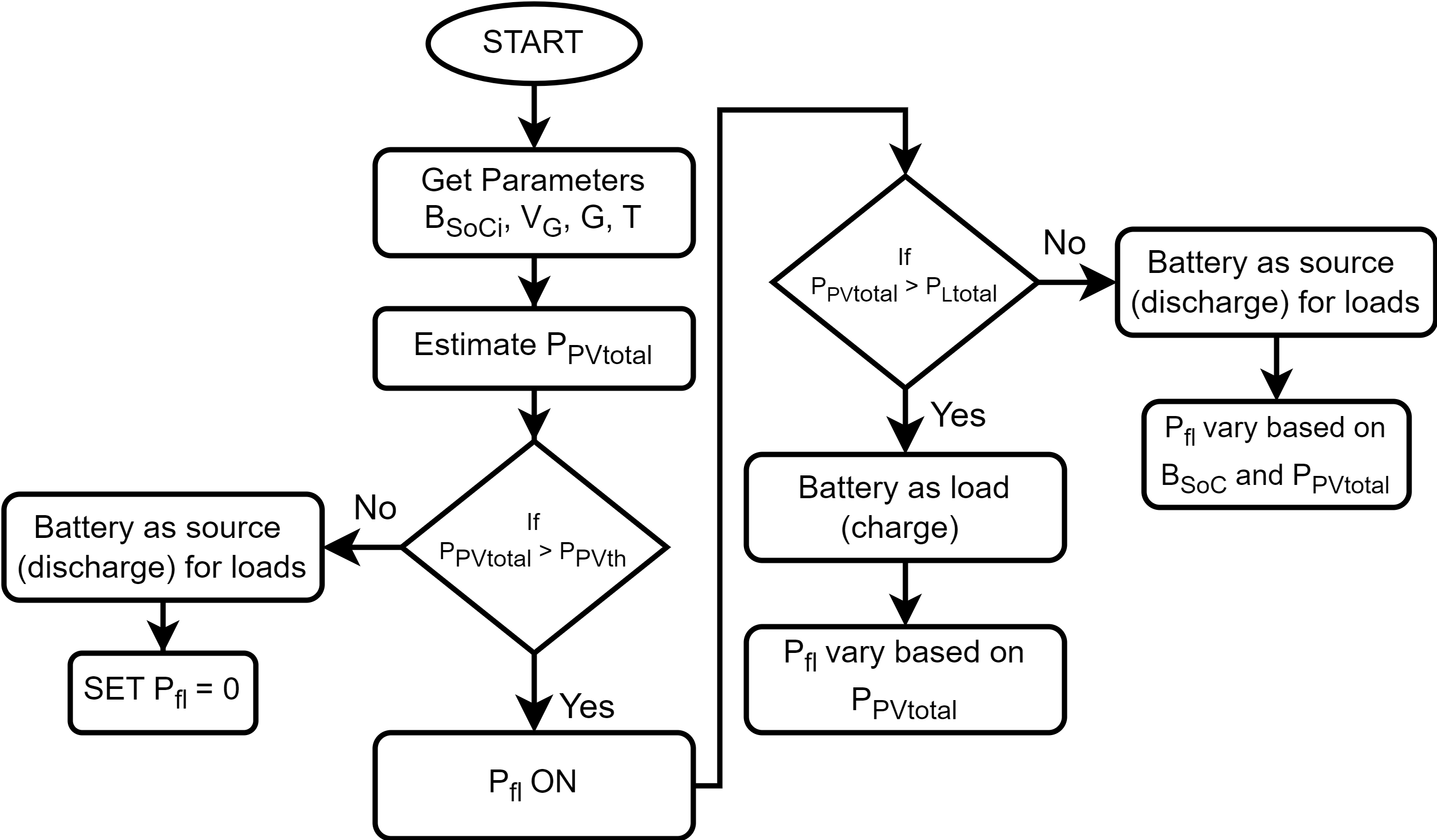}
\caption{Flowchart for the flexible dynamic load actuation.}
\label{fig_7}
\end{figure}

A PI controller is designed as shown in Fig. \ref{fig_6} to control the bidirectional DC-DC converters of the DC microgrid. A flowchart describing the actuation of flexible dynamic load according to the PV power availability is shown in Fig. \ref{fig_7}. The system is designed such that one of the BESS along with the converter will operate in Constant Voltage (CV) mode. Others will act in Constant Current (CC) mode to avoid voltage conflicts in the system. The signals for the status of individual BESS converters for charging and discharging are transferred from Hyphae \cite{15}.

\subsection{Hyphae Software Framework}
\begin{figure}[t]
\centering
\includegraphics[width=3.3in]{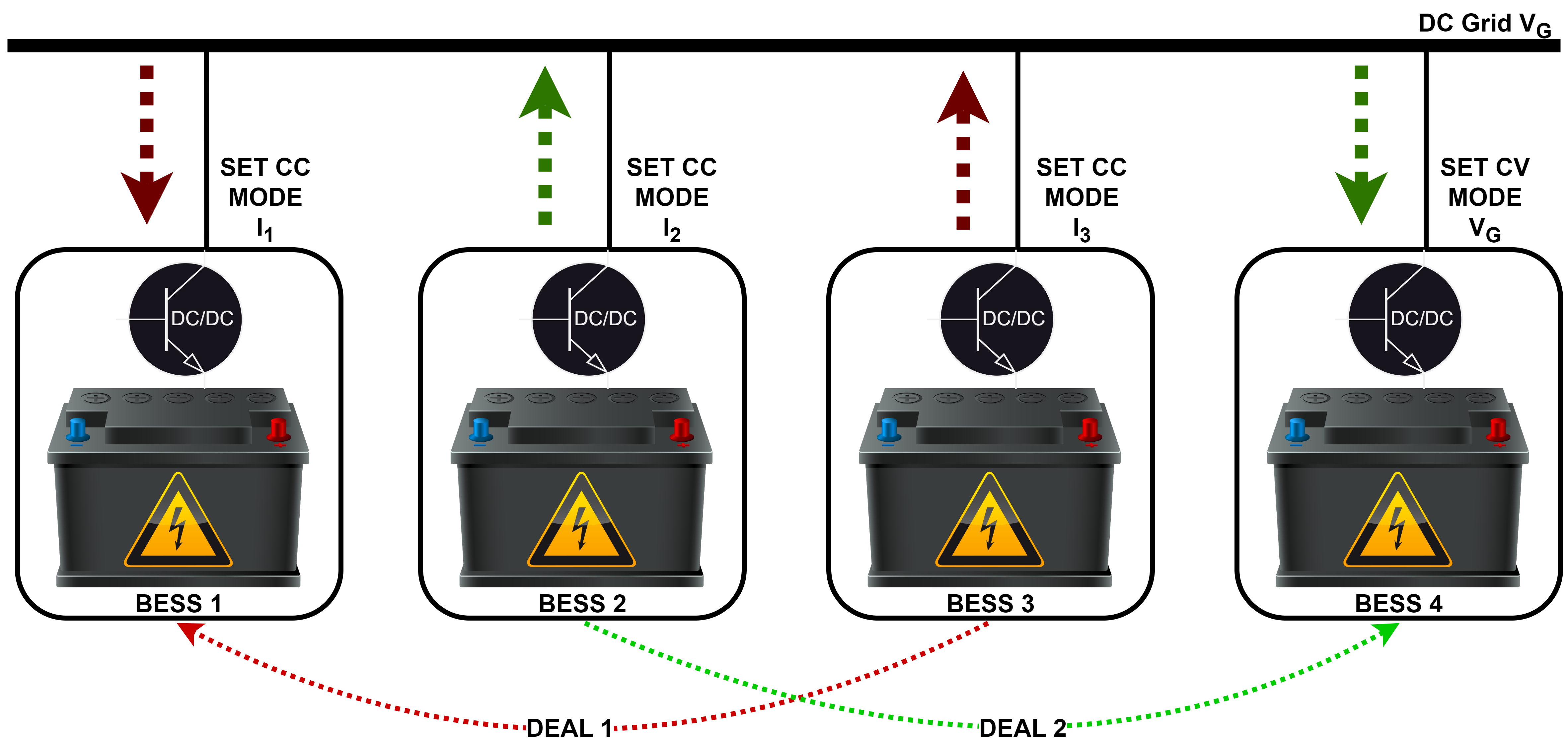}
\caption{Representation of Hyphae framework.}
\label{fig_8}
\end{figure}

The APIS system from the Hyphae decentralized software framework is used to automatically manage the energy transfer between different BESS more efficiently. The status for the individual controllers is given as a CC mode with the respective charge or discharge current with one BESS as CV mode. The constraints for the BESS such as min-max SoC levels, max voltage, and current levels are fed to the Hyphae framework. A two-way information exchange such as SoC, voltage, current, deals, and status between the Hyphae and the grid model is communicated using TCP/IP. The Hyphae framework orchestrates the transactions for the charging and discharging of each BESS, as depicted in Fig. \ref{fig_8}. These transactions are then communicated to the grid system operating on the HIL machine \cite{12}.

\section{Results and Discussion}\label{s:results}
\begin{figure}[t]
\centering
\includegraphics[width=2.2in]{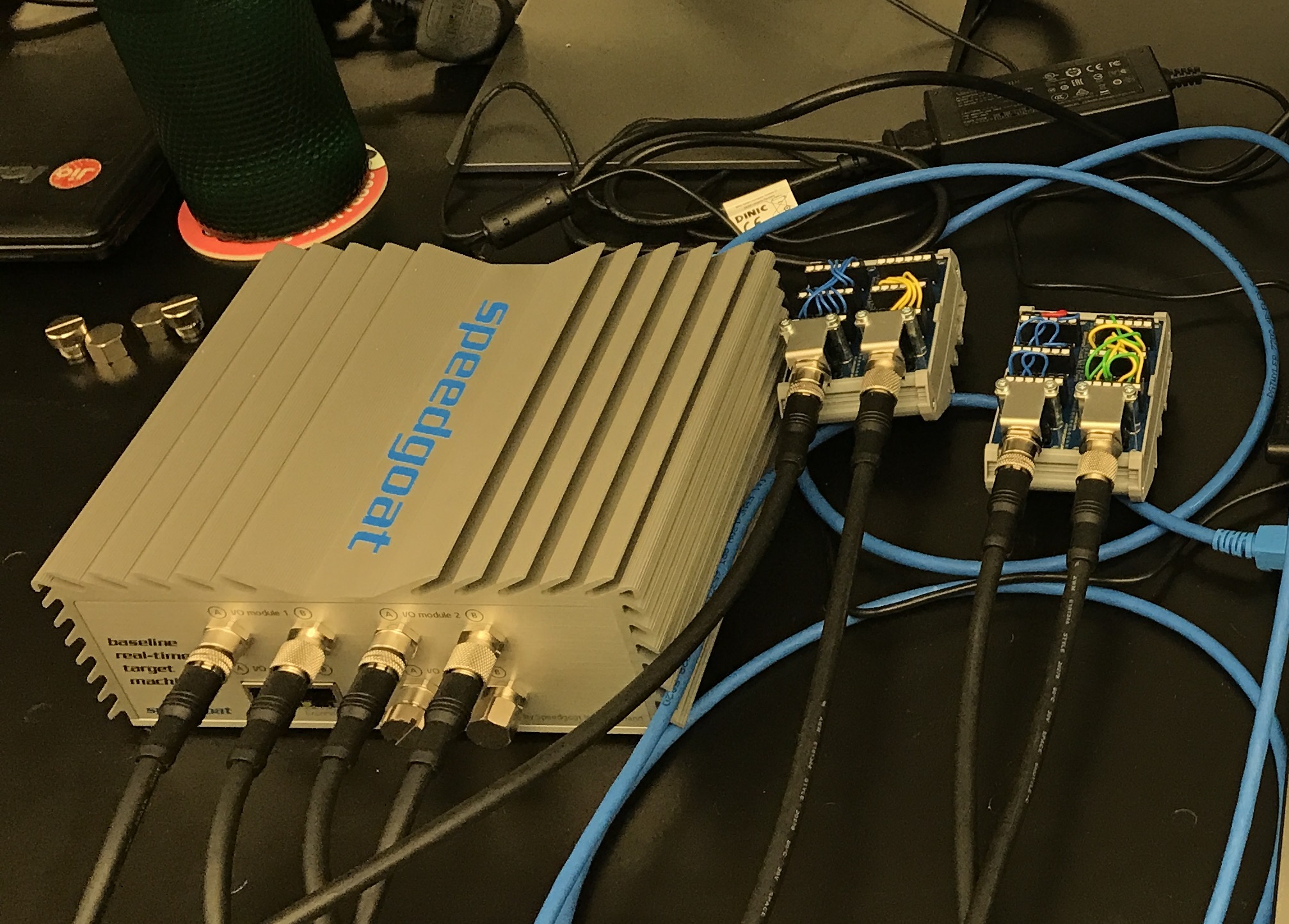}
\caption{Speedgoat baseline real-time target HIL machine.}
\label{fig_9}
\end{figure}

The simulations are performed in the Speedgoat baseline real-time target HIL machine that is shown in Fig. \ref{fig_9}. The results are obtained for different scenarios. The load data is obtained from the New York Independent System Operator (NYISO) for a 24-hr period on a relatively normal day (Friday, February 16, 2024) and an extreme case (Saturday, April 11, 2020) during which, due to the COVID-19 pandemic lockdowns, there were low net-load conditions in the system \cite{lowdem}.

\subsection{Voltage Regulation}

\begin{figure}[t]
\centering
\includegraphics[width=3.6in]{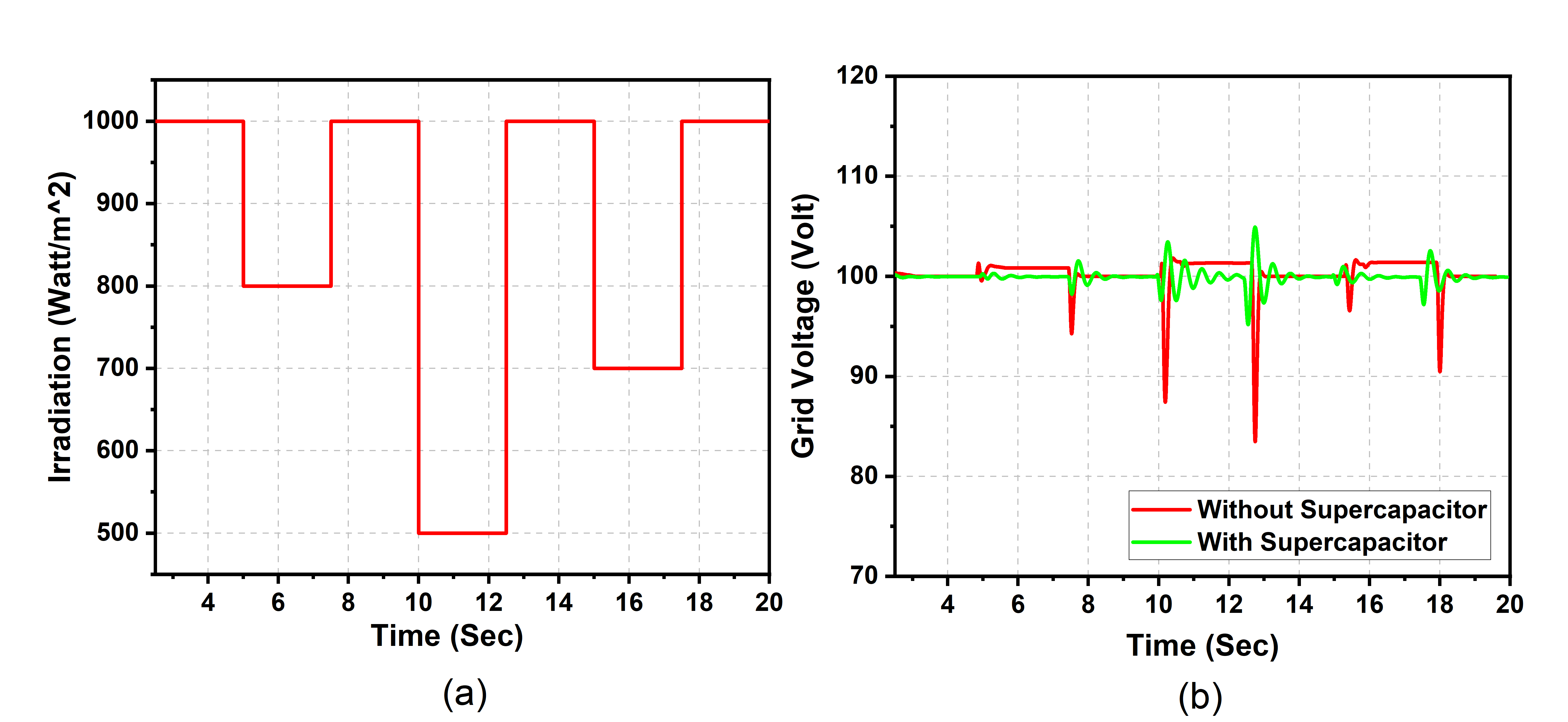}
\caption{(a) Irradiation change, (b) Voltage variation with and without supercapacitor.}
\label{fig_10}
\end{figure}

The voltage variation within the DC microgrid in response to changes in solar PV irradiation is illustrated in Fig. \ref{fig_10}. Results are presented for scenarios with and without the integration of a supercapacitor. A significant drop in grid voltage is observed when irradiation levels change abruptly. However, the inclusion of a supercapacitor markedly mitigates these abrupt voltage fluctuations, enabling the voltage to more closely align with the predetermined setpoint.

\subsection{Flexible Load: Disabled}

\begin{figure}[t]
\centering
\includegraphics[width=2in]{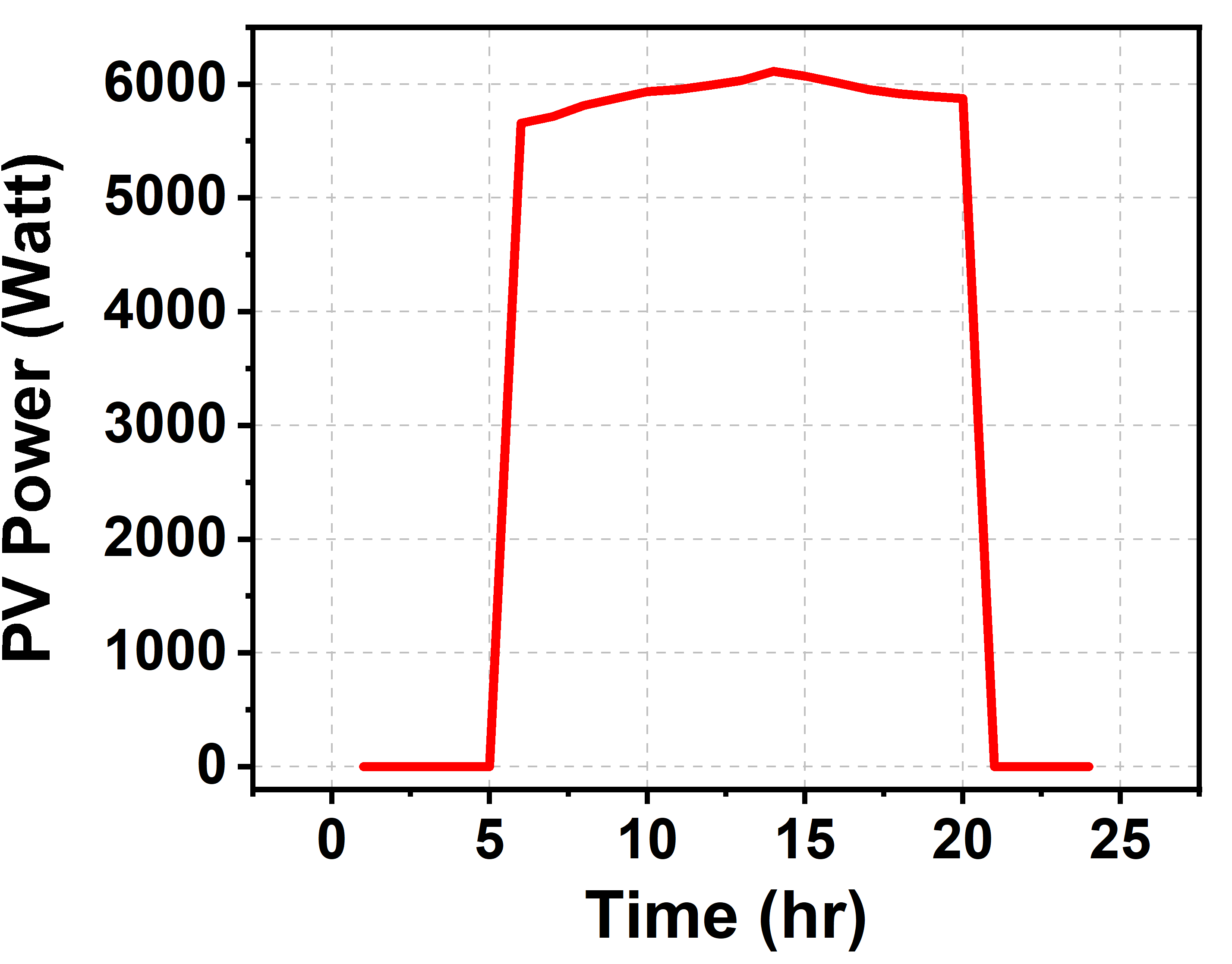}
\caption{{Total PV Power availability for a day}.}
\label{fig_11}
\end{figure}

\begin{figure}[t]
\centering
\includegraphics[width=3.55in]{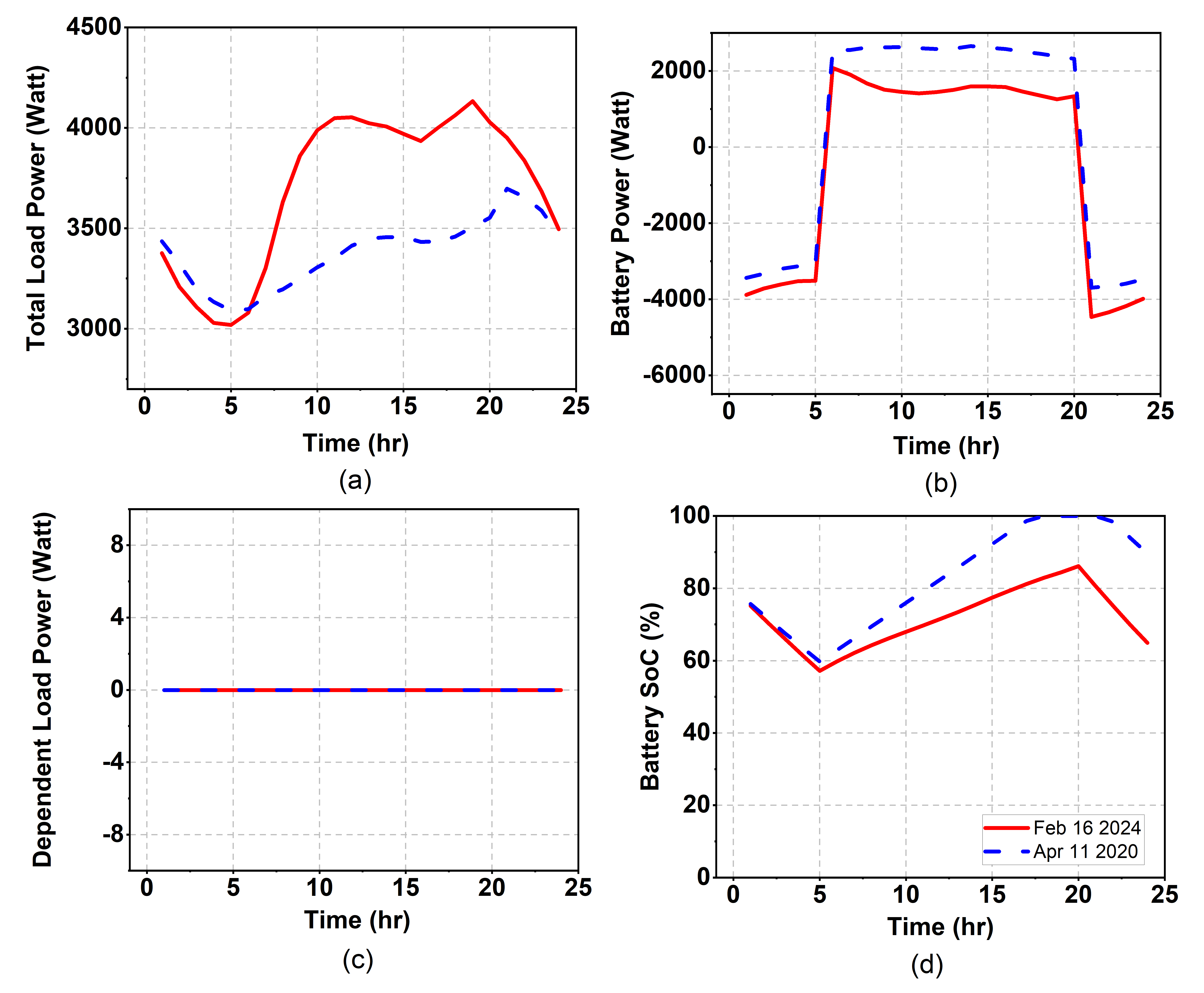}
\caption{{Scenario when dependent loads are off. (a) Total power consumed by all the loads following NYISO data, (b) 
Power delivered from and to the BESS, (c) Flexible load power consumption, and (d) SoC of the battery.}}
\label{fig_12}
\end{figure}

The availability of PV power over a 24-hour period is depicted in Fig. \ref{fig_11}. Fig. \ref{fig_12} presents the results for the power-sharing between BESS and solar PV when the dependent load is inactive. Notably, surplus PV power is utilized to charge the BESS, resulting in an increase in the SoC of the battery. A positive battery power indicates charging, whereas negative power signifies discharging. Upon the completion of the charging cycle of one BESS, another BESS is designated for charging. This power management strategy is orchestrated by the Hyphae system. With the dependent flexible loads deactivated, their power consumption is null.

\subsection{Flexible Dependent Load: Enabled}

\begin{figure}[t]
\centering
\includegraphics[width=3.55in]{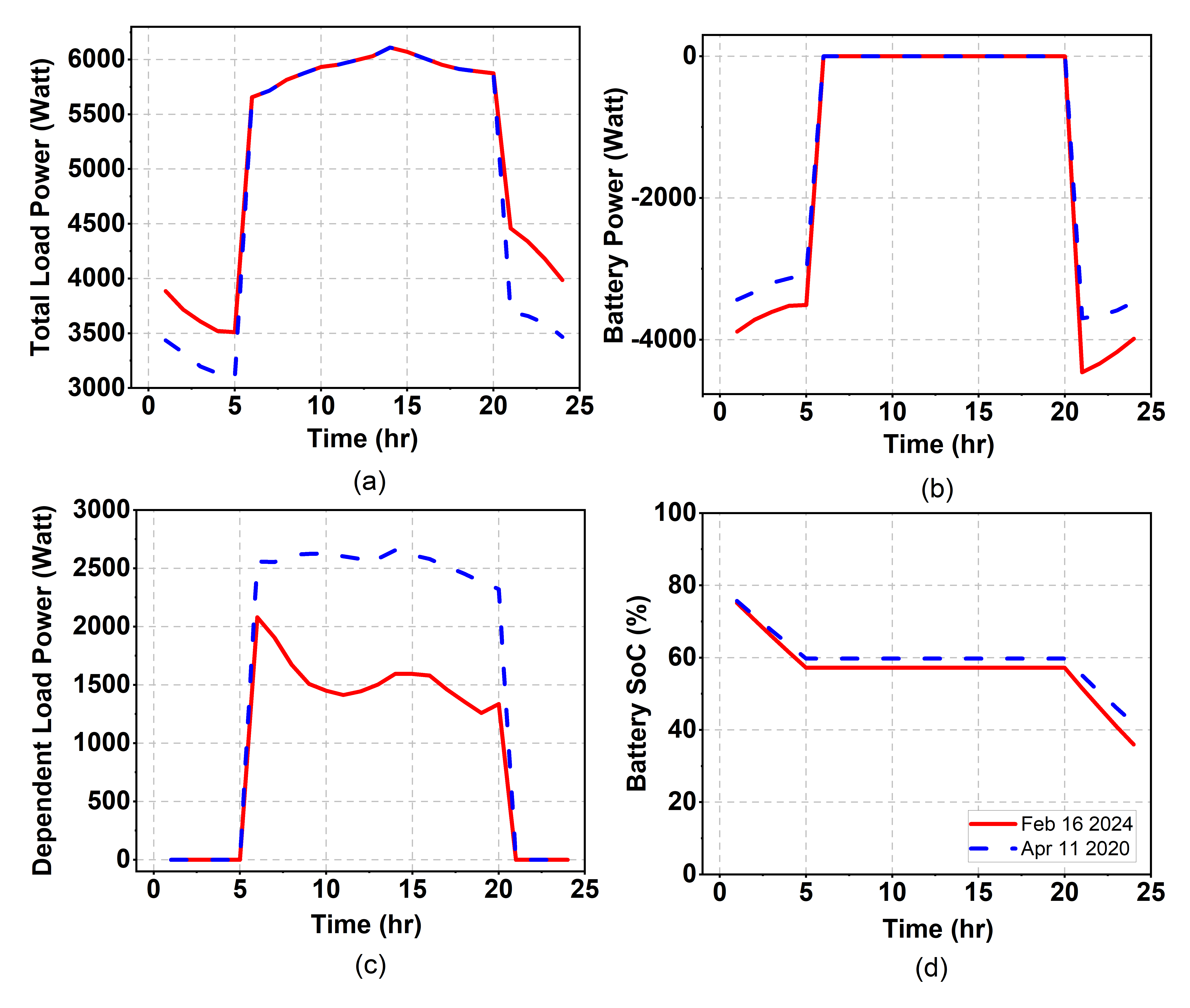}
\caption{{Scenario when dependent loads are fully on. (a) Total power consumed by all the loads, (b) Power delivered from and to the BESS, (c) Dlexible load power consumption, and (d) SoC of the battery.}}
\label{fig_13}
\end{figure}

In this scenario, all surplus PV power is allocated to the dependent flexible loads, bypassing the charging of the BESS. This strategy ensures that the demand of all flexible loads is met in accordance with their requirements and the availability of PV power. Consequently, the power consumption of these dependent flexible loads is directly proportional to the surplus PV power. Therefore, during periods of non-peak hours when excess PV power is available, this energy is utilized to operate dependent flexible loads, such as washing machines and dryers. Fig. \ref{fig_13} presents the results of this configuration.

\subsection{Flexible Dependent Load: Partially Enabled}

\begin{figure}[t]
\centering
\includegraphics[width=3.55in]{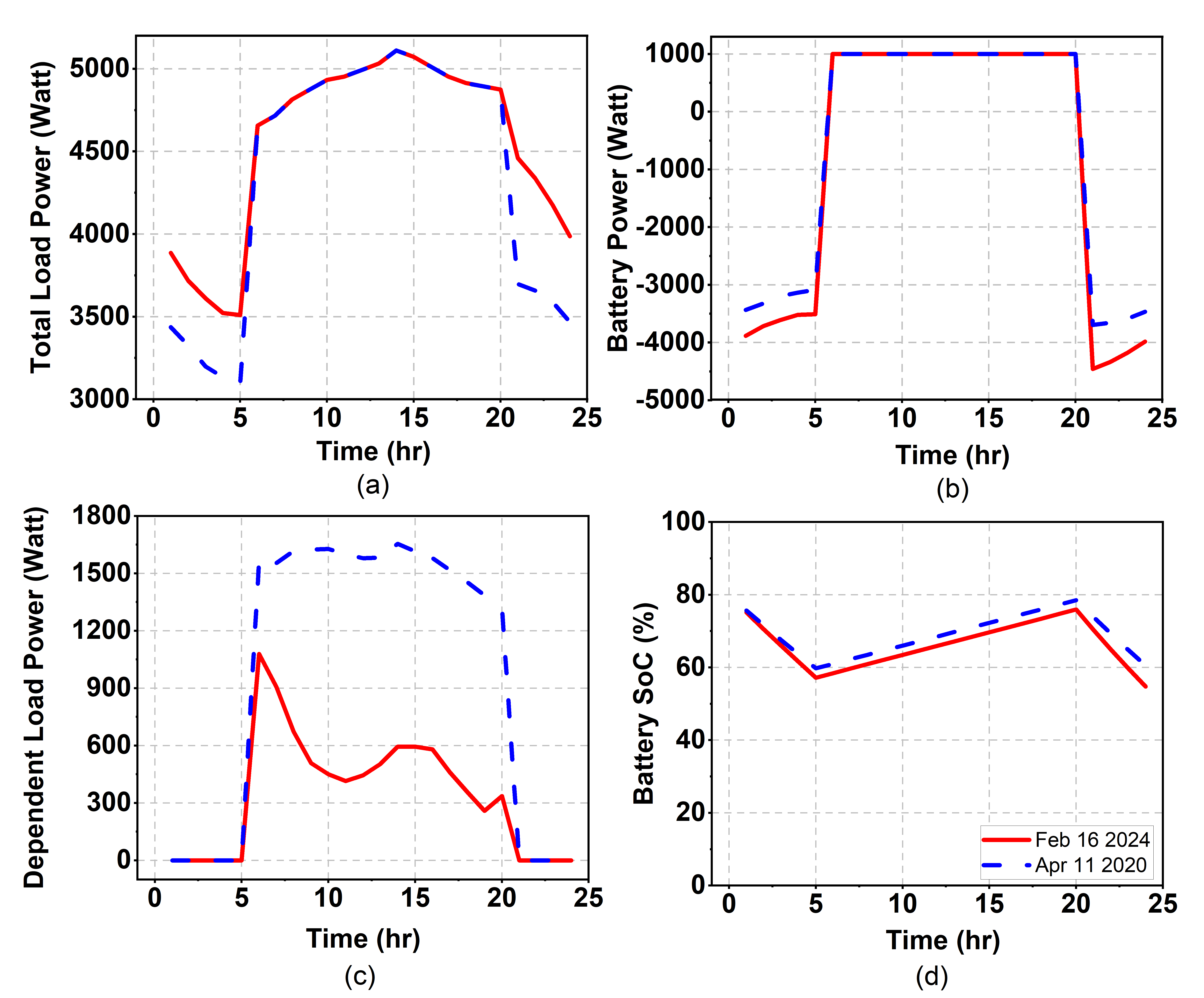}
\caption{{Scenario when dependent loads are partially on. (a) Total power consumed by all the loads, (b) Power delivered from and to the BESS, (c) Flexible load power consumption, and (d) SoC of the battery.}}
\label{fig_14}
\end{figure}

In this scenario, not all surplus PV power is allocated exclusively to the dependent flexible loads; a portion is also directed to the BESS for charging. This approach achieves a balance between partially meeting the demand of flexible loads and ensuring a consistent charging input to the BESS. The previous section outlines how users can set this allocation limit. Fig. \ref{fig_14} displays the results for this case, illustrating an optimized distribution of surplus PV power for enhanced efficiency. Specifically, a part of the excess power from the PV system is utilized to satisfy immediate energy requirements of the flexible dependent loads, thus not relying entirely on BESS for power. Concurrently, in this configuration, the remaining surplus PV power, after catering to the load demand, is diverted to charge the BESS. This strategic distribution serves two purposes: it meets current energy needs while also storing excess energy for future use. The BESS thereby acts as an energy reservoir, accumulating power during times of high solar output to ensure a stable power supply during periods with little to no solar generation, consequently improving the resilience and sustainability of the overall energy system. 

{The results provide us with the overall system behavior in three different modes of operation for different dates. It is demonstrated that the partial enablement of flexible loads in the system serves to their electricity needs along with charging the BESS nodes which can be used during power uncertainties. The communication link between the Hyphae software framework and the modeled DC microgrid setup in the Speedgoat helps the BESS at different nodes to maintain particular SoCs. This whole setup is designed as a highly configurable platform to conduct various studies by defining the behavioral model of each BESS, load patterns, PV forecast, etc.}    

\section{Conclusions}\label{s:conclusions}

This work introduces a comprehensive DC microgrid testbed, incorporating a Speedgoat real-time HIL machine and integrating various DERs alongside non-flexible and flexible loads. The testbed's design is modular, facilitating easy integration with additional power-generating and consuming entities. Energy exchanges between the BESS are automated through the Hyphae framework, while voltage regulation within the grid is achieved using a supercapacitor. To mitigate the uncertainties associated with PV power generation, a dynamically adjustable flexible load component is introduced. The investigation encompasses three distinct scenarios: with flexible loads disabled, enabled, and partially enabled. Results indicate that incorporating flexible loads into the grid maximizes the utilization of available PV power, preventing losses. This testbed provides a valuable platform for simulating various load conditions and control strategies, offering insights for further research and system optimization.

\section*{Acknowledgment}
This publication is based upon work supported by King Abdullah University of Science and Technology (KAUST) under Award No. RFS-OFP2023-5505.

\bibliographystyle{IEEEtran}
\bibliography{RefDCG}
\end{document}